\documentclass{article}

\usepackage{arxiv}
\usepackage{algorithm}
\usepackage{algorithmic}
\usepackage{multirow}
\usepackage[utf8]{inputenc} 
\usepackage[T1]{fontenc}    
\usepackage{hyperref}       
\usepackage{url}            
\usepackage{booktabs}       
\usepackage{amsfonts}       
\usepackage{nicefrac}       
\usepackage{microtype}      
\usepackage{lipsum}		
\usepackage{graphicx}
\usepackage[square, numbers, comma, sort&compress]{natbib}%
\hypersetup{urlcolor=blue, colorlinks=true}
\usepackage{backref}

\usepackage{amsmath}
\usepackage{doi}

\newcommand{\algoname}{AsynCoMARL }
\newcommand{\algonameNoSpace}{AsynCoMARL}

\title{Asynchronous Cooperative Multi-Agent Reinforcement Learning with Limited Communication}

\author{ Sydney Dolan
    \\
	Department of Aeronautics and Astronautics\\
	Massachusetts Institute of Technology\\
	Cambridge, MA 02139 \\
	\texttt{sydneyd@mit.edu} \\
	\And
	Siddharth Nayak \\
	Department of Aeronautics and Astronautics\\
	Massachusetts Insitute of Technology\\
	Cambridge, MA 02139 \\
	\texttt{sidnayak@mit.edu} \\
    \And
	Jasmine Jerry Aloor \\
	Department of Aeronautics and Astronautics\\
	Massachusetts Institute of Technology\\
	Cambridge, MA  \\
	\texttt{jjaloor@mit.edu} \\
    \And
     Hamsa Balakrishnan \\
	Department of Aeronautics and Astronautics\\
	Massachusetts Institute of Technology\\
	Cambridge, MA  \\
	\texttt{hamsa@mit.edu} \\
}

\date{}

\renewcommand{\shorttitle}{AsynCoMARL}

\hypersetup{
pdftitle={A template for the arxiv style},
pdfsubject={q-bio.NC, q-bio.QM},
pdfauthor={David S.~Hippocampus, Elias D.~Striatum},
pdfkeywords={First keyword, Second keyword, More},
}

\begin{document}
\maketitle

\begin{abstract}
	We consider the problem setting in which multiple autonomous agents must cooperatively navigate and perform tasks in an unknown, communication-constrained environment.  Traditional multi-agent reinforcement learning (MARL) approaches assume synchronous communications and perform poorly in such environments. We propose AsynCoMARL, an asynchronous MARL approach that uses graph transformers to learn communication protocols from dynamic graphs. AsynCoMARL can accommodate infrequent and asynchronous communications between agents, with edges of the graph only forming when agents communicate with each other. We show that AsynCoMARL achieves similar success and collision rates as leading baselines, despite 26\% fewer messages being passed between agents.  
\end{abstract}

\section{Introduction}

Communication is crucial in cooperative multi-agent systems with partial observability, as it enables a better understanding of the environment and improves coordination. In extreme environments such as those underwater or in space, the frequency of communication between agents is often limited \cite{space,marine}. For example, a satellite may not be able to reliably receive and react to messages from other satellites synchronously due to limited onboard power and communication delays. In these scenarios, agents aim to establish a communication protocol that allows them to operate independently while still receiving sufficient information to effectively coordinate with nearby agents. 

Multi-agent reinforcement learning (MARL) has emerged as a popular approach for addressing cooperative navigation challenges involving multiple agents. Reinforcement learning approaches learn policies directly by interacting with a simulated environment for policy improvement. Unlike planning-based solutions \cite{coord_exp}, which require non-trivial implementation heuristics and expensive inference computation at execution time, reinforcement learning offers a way to represent complex strategies with minimal overhead once the policies are trained. The classical MARL formulation follows a synchronous framework, where all agents take actions simultaneously, with actions executed immediately at each time step. Similarly, communications between agents [are assumed to] occur instantaneously, frequently, and synchronously, often with agents broadcasting their state to every other agent in the environment. As a result, traditional MARL algorithms are poorly suited to asynchronous settings where agents operate on independent time scales and cannot frequently communicate with one another. While prior work has explored how to coordinate agents using less communication, differences in message encoding approach can significantly impact performance, as shown in Section \ref{exp_result}. Furthermore, little attention has been given to achieving such coordination asynchronously. Unfortunately, existing asynchronous approaches often increase communication to compensate for the lack of synchronized coordination among agents. Our work builds on prior work in multi-agent communication by introducing an asynchronous MARL framework that enables agents to minimize communication while completing navigation tasks. The main contributions of this paper are:
\begin{enumerate}
    \item We propose \algonameNoSpace, a graph transformer-based communication protocol for MARL that relies on dynamic graphs to capture asynchronous and infrequent communications between agents.
    \item We empirically evaluate \algoname on two MARL benchmarks (Cooperative Navigation \cite{dolan2023satellite} and Rover-Tower \cite{aac}) and show that our method can achieve superior performance while using less communication. 
\end{enumerate}

\algoname is an asynchronous MARL approach that leverages graph transformers to learn communication protocols from dynamic graphs. In this setting, agents seek to learn the communication protocol that best utilizes asynchronous and infrequent communications from nearby agents. Each agent's graph transformer utilizes a dynamic weighted directed graph to learn a communication protocol with other active agents in its vicinity. The underlying MARL algorithm uses this learned graph transformer encoding in both the actor and the critic to optimize action selection, leading to effective cooperation with less communication between agents. We conduct experiments in two environments, Cooperative Navigation and Rover-Tower, chosen to replicate the communication-constrained settings of space missions and planetary rover exploration. We find that strategies learned by \algoname use less communication and outperform other MARL methods.

\section{Related Work}

\subsection{Attention and Graph-Based Methods for Multi-Agent Communication} 
Through communication, agents can obtain a better understanding of their own environment and other agents' behaviors, thus improving their ability to coordinate. For a comprehensive survey on research on the impact of communication in multi-agent collaboration, we refer the reader to \cite{comm_survey}. Seminal works such as CommNet \cite{CommNet} used a shared neural network to process local observations for each agent. Each agent makes decisions based on its observations and a mean vector of messages from other agents. Although agents can operate independently with copies of the shared network, instantaneous communication with all agents is necessary. Subsequent works \cite{IC3Net, ATOC} aimed to investigate methodologies that improved coordination by using a small subset of agents. In ATOC \cite{ATOC}, the authors used a probabilistic gate mechanism to communicate with nearby agents in an observable field and a bi-LSTM to concatenate messages together. Further improvement to multi-agent collaboration was made by learning encodings of local information via attention mechanisms. TarMAC \cite{TarMAC} uses an attention mechanism to generate encodings of the content of messages so that nearby agents can learn the message's significance via a neural network implicitly. While attention-based mechanisms can help agents improve their abilities to utilize individual pieces of communication, attention-based approaches neglect larger inter-agent relationships. Graph-based methods like DGN~\cite{DGN}, DICG~\cite{DICG}, InforMARL \cite{informarl_icml}, MAGIC \cite{magic}, and EMP~\cite{EMP} formulate interactions between different agents as a graph. These methods use graph structures to learn relational representations from nearby nodes. However, a similar communication problem to CommNet arises with these methodologies, as they often require fully connected graphs during the learning process.  For example, in EMP, all agents must know the position of all other entities in the graph at the beginning of an episode. This assumption is a key limitation, as it requires that all agents will be able to coordinate synchronously.

 \subsection{Event-Triggered Communication} 
 Event-triggered control is a control strategy in which the system updates its control actions only when certain events or conditions occur, rather than continuously or at regular time intervals \cite{ETC_descript}. By reducing the number of control updates, event-triggered control can significantly decrease the workload on controllers, a beneficial attribute for systems with limited resources. ETC~\cite{ETC}, VBC~\cite{VBC}, and MBC~\cite{MBC} have proposed event-triggered control methodologies to reduce the communication frequency and address communication constraints. These works focus on optimizing when transmission can occur rather than optimizing how to achieve better performance with less communication. \citet{Menda_2019} frame the asynchronous decision-making problem where agents choose actions when prompted by an event or some set of events occurring in the environment. By relying on a continuous state-space representation and an event-driven simulator, the agents step from event to event, and lower-level timestep simulations are eliminated. This improves the scalability of the algorithm but does not capture low-level agent-agent interactions during an event. 
\subsection{Asynchronous Actor Critic} 
An alternative approach to asynchronous environments formulates the problem as a Macro-Action Decentralized Partially Observable Markov Decision Process (MacDec-POMDP) \cite{macpomdp}, where agents can start and end higher level 'macro' actions at different time steps. \citet{IAICC} propose a methodology for multi-agent policy gradients that allow agents to asynchronously learn high-level policies over a set of pre-defined macro actions.  The method relies on an independent actor independent centralized critic training framework, named IAICC, to train each agent. \citet{AOCC}  alter the IAICC framework to encode agent time history independently to avoid duplicate macro-observations in the centralized critic. These works focus on learning algorithms for asynchronous\textit{ macro-actions} that occur across multiple time steps. By contrast, our work focuses on the micro level, looking for planning opportunities while agents achieve the same macro-level task.

\begin{figure*}[t]
    
    
    \includegraphics[width=1.0\textwidth]{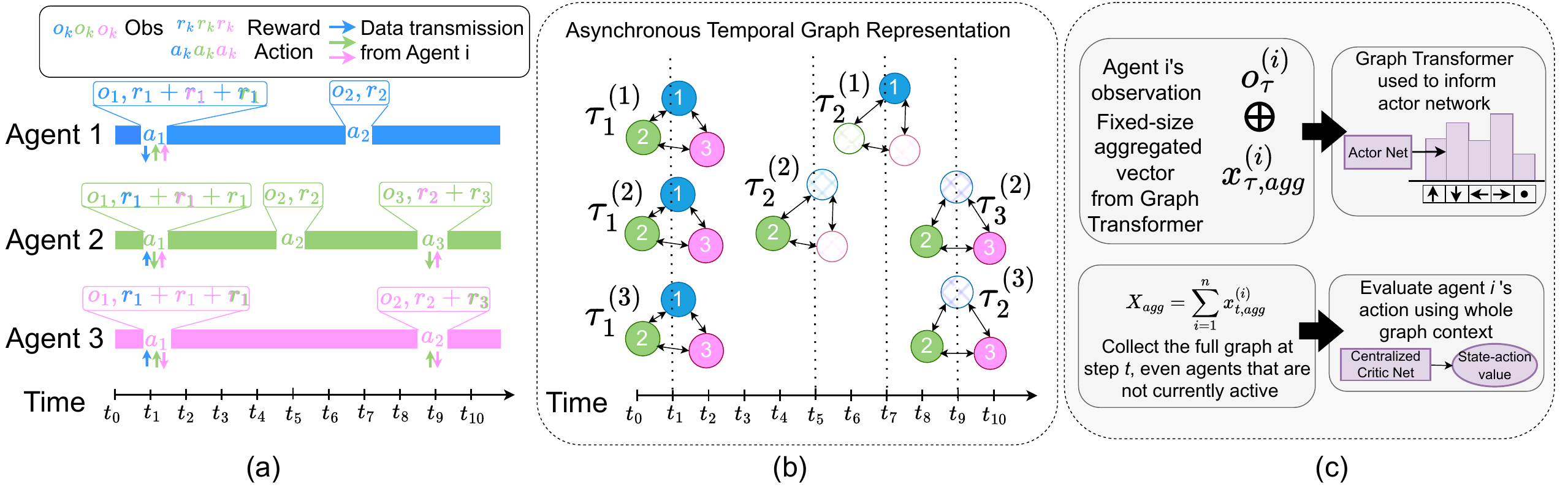}
    \caption{Overview of \algonameNoSpace: (a) Environment. Agents within our environment take actions and observations asynchronously. To encourage collaboration, when agents take actions at the same time $t$, they receive a shared reward. The sequence of actions and observations for agent $i$ is referred to by timescale $\tau^{(i)}$. The arrows indicate data transmissions, which represent the most recent graph observation $x_{\tau_{agg}}^{(i)}. $ b) Asynchronous Temporal Graph Representation. Each active agent within our environment is translated to become a node on the graph, and they can communicate with other agents located nearby within distance $\phi$. Our graph representation is dynamic, meaning that graph edges connect and disconnect depending on agent proximity.  c) Agent $i$’s observation is combined with its node observations from the graph transformer, $x_{\tau,agg}^{(i)}$, and fed into the actor network. The critic takes the full graph representation $X_{agg}$ and evaluates agent $i$’s action.}
    \label{alg_over}
\end{figure*}
\section{Preliminaries}
\subsection{Decentralized Partially Observable Markov Games} 

A decentralized partially observable Markov decision process (DEC-POMDP) is a multi-agent extension of a Markov decision process (MDP) where multiple players (agents) interact in a shared environment, but each agent has only limited or partial information about the current state of the environment and the state of other agents. A DEC-POMDP for $\mathcal{N}$ agents can be defined by a set of global states $\mathcal{S}$, a set of private observations for each agent $o^{(1)},o^{(2)},..,o^{(n)}$, a set of actions for each agent,$a^{(1)},a^{(2)},..,a^{(n)}$,  and the transition function $T: \mathcal{S} \times a^{(1)}\times .. \times a^{(N)} \rightarrow \mathcal{S}$. Agent $i$ chooses actions $a^{(i)} \in \mathcal{A}^{(i)}$, and then obtains a reward as a function of its state-action pairing: $r^{(i)}: \mathcal{S} \times a^{(i)} \rightarrow \mathbb{R}$. It also receives a local observation $o^{(i)}$.  Agent $i$ aims to maximize its reward $R^{(i)}=\sum_{t=0}^T r^{(i)}$.

\subsection{Policy Gradient Methods}
The policy gradient method is used in RL tasks to perform gradient ascent on the agent policy parameters, $\theta$, to optimize the total discounted reward, $J(\theta) =\mathbb{E}_{s\sim\rho^\pi, a\sim\pi_\theta}[R]$. Here, $\rho^{\pi}$ is the state distribution, $\pi_\theta$ is the policy distribution and $R$ represents the reward at that time $t$, $R_t=\sum_{t'=t}^T r(s_{t'},a_{t'})$.
\begin{equation}
    \nabla_\theta J(\theta) = \mathbb{E}_{s\sim\rho^\pi, a\sim\pi_\theta} [\sum_{t=1}^T \nabla_\theta \log \pi_\theta(a_t |s_t) R_t]
\end{equation}
In lieu of $R_t$, we use the advantage function $A^\pi(s_t,a_t)= Q^{\pi}(s_t,a_t)-V^{\pi}(s_t)$ to decrease the variance of the estimated policy gradient, where $V^\pi(s_t)$ is the value function. 

We adopt a centralized training decentralized execution learning paradigm \cite{MADDPG, MAPPO}. In execution, agents learned policies are conditioned only on their own action-observation history. In training, the critic has access to the global state $\mathcal{S}$ of the environment, which is used to train the model end to end. 

\section{Methodology}
In this section, we introduce our proposed multi-agent graph transformer communication protocol, \algonameNoSpace. We consider a partially observable setting of $N$ agents, where agent $i$ receives local observation $o_i^t$ at time $t$, containing local information from the global state $\mathcal{S}$. The agent, $i$, learns a communication-based policy $\pi_i$ to output a distribution over actions $a_t^{(i)}$ $\sim$ $\pi^{(i)}$ at time step $t$. Here, we present a description of our graph formulation and its integration into the graph transformer, a description of our framework's key components, and our training procedure. 
\subsection{Asynchronous Formulation}
\label{action-communciation}

In traditional multi-agent reinforcement learning, agents always take action steps synchronously without considering the communication constraints associated with their action selection. For instance, when exploring unknown environments, a single rover may navigate to an area where they are unable to communicate with others, or the duration of transmitting a single message may take several time steps. To model these communication-action costs, we define a new time scale $\tau$ to account for the specific actions each agent has taken at different time steps.  As shown in Figure \ref{alg_over}, $\tau_1^{(i)}$ represents the time of the first action that agent $i$ has taken. In Figure \ref{alg_over} panel (b), all three agents take their first action at the same $t$, resulting in the same time reference point for $\tau_1$. However, the next time $t$ that each agent takes their next action is different (agent 1 $\tau_2^{(1)}=t_7$, agent 2 $\tau_2^{(2)}=t_5$ agent 3 $\tau_2^{(3)}=t_9$). Rather than incorporating all time steps into the replay buffer, we only include those steps where the agents are taking an action. As a result, the replay buffer of each agent's trajectory is based on their $\tau$ sequence of actions instead of $t$. To improve the generalizability of our algorithm to different periods of time between actions $\tau_1$ and $\tau_2$, during training, we randomly generate the period between different actions.  We refer to this randomized parameter as $\mu$, and it determines when an agent will take a subsequent action. $\mu$ is chosen from a uniform distribution, with different intervals used for training and testing to ensure diversity. The $\mu$ range is kept relatively small to model asynchronous behavior that maintains staggered interactions between agents without diverging to entirely independent timescales. For each initiation, the delay is selected randomly from this distribution and remains constant throughout the episode rather than changing after specific actions. This decision avoids creating an artificial link between certain movement actions and specific delays, as we consider such realism beyond the scope of this work and more applicable to domain-specific problems.

\subsection{Graph Overview}
 Graph transformers are a specialized type of transformer model designed to handle graph-structured data. In addition to the traditional transformer's ability to capture long-range dependencies, graph transformers enable the model to understand both local node interactions via graph edges and global context via self-attention. 

 As an input into our graph transformer, we form an agent-entity graph~\cite{EMP}, where every object in the environment is assumed to be either an agent, an obstacle, or a goal. All objects within the environment are transformed to be nodes on the graph, with node features $x_j=[p^j_i, v^j_i, p^{\mathrm{goal},j}_i, \texttt{entity\_type(j)}]$ where $p^j_i, v^j_i, p^{\mathrm{goal},j}_i$ are the \emph{relative} position, velocity, and position of the goal of the object at node $j$ with respect to agent $i$, respectively. If node $j$ corresponds to a (static/dynamic) obstacle or a goal, we set $p^{\mathrm{goal},j}_i \equiv p^j_i$. To process the \texttt{entity\_type} categorical variable, we use an embedding layer.

 Our graph formulation is dynamic, meaning that edges are formed depending on an entity's proximity to other objects in the graph and their corresponding communication interval, as shown by $(b)$ in Figure \ref{alg_over}. We define an edge to exist  $e\in\mathcal{E}$ between an agent and another agent if they are within a `communication radius' $\lambda$ of each other and if they are taking an action at the same time step $t$. Edges are formed between agents and obstacles or landmarks when they are merely within the agent's communication radius. As landmarks and obstacles do not have decision-making abilities, their presence can be sensed at every time step the agent is in proximity. When an edge is formed, $e_{ij}$, it has an associated edge feature given by the Euclidean distance between the entities involved, $i$ and $j$.

For a complete representation of our graph structure, we create an adjacency matrix to represent the connectivity between different nodes on a graph. An adjacency matrix $\mathbb{A}$ is a square matrix used to describe the connections between nodes on a graph, where each element $\mathbb{A}_{i,j}$ represents the presence or absence of an edge between node $i$ and node $j$. Our graph is directed and weighted, so the value in the adjacency matrix for $\mathbb{A}_{i,j}$ is proportionate to the weight of the edge between nodes $i$ and $j$. To account for the fact that our agents can take actions at different time steps, we first introduce the variable $d_i$ to represent the current status of the agent, where $d_i=1$ if the agent is active and $d_i=0$ if the agent is inactive. We create an additional matrix to reflect the activity status of all nodes in the graph $\mathcal{D} \in \mathbb{R}^{N\times N}$ where $\mathcal{D}[i,j] = d_i\cdot d_j$, where $d_i$ and $d_j$ reflect whether node $i$ and node $j$ are active. This is used to ensure that interactions can only occur if both nodes are active. The masked adjacency matrix $\mathbb{A}^{\textrm{masked}}$ is given by the element-wise product  $\mathbb{A} \circ \mathcal{D}$. The resulting masked adjacency matrix dynamically updates as agents become inactive, allowing the algorithm to focus on interactions between remaining agents. Similarly, a death mask enforces that no interactions or learning updates are computed for agents whose tasks are already completed.

\subsection{Graph Transformer}
We rely on a graph transformer to encode messages and characterize relationships between different entities in the environment. Our transformer takes the node features $x_k$ and edge features $e_{k}$ as input. The graph transformer is based on a Unified Message Passing Model (UniMP) \cite{graphTransformerConv} that relies on multi-head dot product attention to selectively prioritize incoming messages from their neighbors based on their relevance. We rely on two layers of this model, with each layer update defined as \begin{equation}
     x_i' = W_1 \cdot x_i + \sum\limits_{j\in\mathbb{N}(i)}\alpha_{i,j}W_2 \cdot x_j
 \end{equation} where $x_k$ are the node features in the graph, $\mathbb{N}(i)$ is the set of nodes which are connected to node $i$, $W_k$ are learnable weight matrices and the attention coefficients. $\alpha_{i,j}$ is computed via multi-head dot product attention

\subsection{Reward Structure} 
At every time step, each agent gets a distance-based reward to an assigned goal, ${\mathcal{R}}_\mathrm{dist}(s_{\tau}, a^{(i)}_\tau)$. When an individual agent, $i$, reaches their assigned goal (indicated by $\rho$), it receives a goal-reaching reward $\mathcal{R}_\mathrm{goal}(s_\tau, a^{(i)}_\tau)$. $\rho$ is 1 if the agent reached the assigned goal and was previously not at the assigned goal; otherwise, 0.  We also penalize agents colliding with other agents or obstacles in the environment using a collision penalty $-C$. $\kappa$ is a 0/1 variable that indicates if an agent collided with another agent or an obstacle.
The reward for agent $i$ at time step $t$ then becomes 
\begin{equation}
    \mathcal{R}_\mathrm{\tau}^{(i)}(s^{(i)}_\tau, a^{(i)}_\tau) = {\mathcal{R}}_\mathrm{dist}(s^{(i)}_\tau, a^{(i)}_\tau) + \rho \mathcal{R}_\mathrm{goal}(s^{(i)}_\tau, a^{(i)}_\tau) - \kappa C
\label{eq:reward}
\end{equation}

Rewards are shared between all agents active at step $t$. The purpose of sharing rewards between active agents is to encourage synchronous collaboration when possible.

\begin{equation}
    \mathcal{R}_\mathrm{total}(s_\tau, A_\tau) = \sum_{i}^{\eta_{act}} \mathcal{R}_\mathrm{\tau}^{(i)}(s^{(i)}_\tau, a^{(i)}_\tau) 
\label{eq:totalreward}
\end{equation}

\subsection{Training}

For our algorithm, the traditional DEC-POMDP setup is augmented by the existence of the graph network, transforming the tuple that describes the problem to: $ \langle \mathcal{N}, \mathcal{S}, \mathcal{O},\mathcal{A}, R,P,\gamma, \mathcal{G} \rangle$. $g^{(i)} = \mathcal{G}(s; i)$ represents the graph network formed by the entities of the environment with respect to agent $i$. Features of the graph structure are used in the policy gradient learning architecture. 

Our learning architecture relies on a centralized critic, which uses the state-action pairs and information from the graph formulation. As shown in Figure \ref{alg_over},  the critic receives the full graph embeddings via a global mean pooling operator ($X_{\mathrm{agg}} = \frac{1}{N}\sum\limits_{i=1}^N x^{(i)}_{\tau,\mathrm{agg}}$) so that it has a global view of the graph representation. The critic updates parameter $\phi$ to approximate the value function.
\begin{equation}
V_{\phi}^{\pi_{\theta_i}}(s_\tau^{(i)},a_\tau^{(i)},X_{agg})= \mathbb{E}_{a \sim \pi_{\theta_i}} \left[ \sum_{\tau=0}^T \gamma^\tau \mathcal{R}_\tau |s_0 =s\right]
\end{equation}

We then create an individual actor for each agent. Each agent relies on policy  
$\pi_{\theta_i}^{(i)}(a_\tau^{(i)}| o_\tau^{(i)},g_\tau^{(i)})$,  parameterized by $\theta_i$ to determine its action $a^{(i)}$ from its local observation $o^{(i)}$ and its local graph network $g^{(i)}$. The resultant policy gradient is:

\begin{equation}
    \nabla_{\theta_i}J(\theta_i) = \mathbb{E}_{{\pi}_{\theta_i}}[\nabla_{\theta_i} \log \pi_{\theta_i}(a_\tau^{(i)}| o_\tau^{(i)},g_\tau^{(i)}) \cdot  V_{\phi}^{\pi_{\theta_i}} (s_\tau^{(i)},a_\tau^{(i)},X_{agg} ) ]
\end{equation}

Algorithm \ref{ACMTrainingSetup} shows the complete training setup for \algonameNoSpace. As mentioned in Section \ref{action-communciation}, each agent $i$ has its own interval $\mu$, which determines when it will take a subsequent action. The interval is reset for each environment scenario to avoid overfitting. During a training episode, an agent's status is marked as active if there have been $\mu$ steps since its last action. If an agent reaches its goal before the end of the episode, its status $d_i$ is changed to 0, thus masking its presence in the masked adjacency matrix $A^{\textrm{masked}}$ passed into the centralized critic function.

\begin{algorithm}[t]
\caption{\algoname Training setup}\label{alg:MBH}
\begin{algorithmic}[1]
\STATE Initialize a decentralized policy network for each agent $i$: $\vec{\pi}_{\theta_i}$
\STATE Initialize a centralized critic network $V_\phi^{\vec{\pi}_\theta}$
\FOR{$episode=1:M$}
\STATE Reset the environment
\STATE Reset buffer $\mathcal{D} = $ \{\}
\FOR{step $t=0:T_f$}
\IF { $\exists i \in \{ 1,..,n \}$ $\text{agent}^{(i)}$.status = active}
\STATE $a^{(i)} $ $\leftarrow $get actions of active agents
\STATE Update environment state with $a^{(i)}$
\STATE $\mathcal{R}_{total}^{(i)} \leftarrow \sum_{j=1}^{\eta_{active}} \mathcal{R}_t^{(j)}$
\STATE Collect $\langle$$o^{(i)}, a^{(i)}, g^{(i)}, r^{(i)}$$\rangle $ into buffer $\mathcal{D}$ at $\tau^{(i)}$
\ELSE
\FOR {$ i \in \{ 1,..,n \}$} \STATE $a^{(i)}$ = $\emptyset$
\STATE Update environment state with $a^{(i)} $
\STATE $\text{mask}(i)  = 
 \begin{cases} 
1, & \text{if agent } i \text{ is finished} \\ 
0, & \text{otherwise}
\end{cases}$
\ENDFOR

\ENDIF
\ENDFOR
\STATE Follow standard MARL update process from MAPPO \cite{PPO}

\ENDFOR
\end{algorithmic}
\label{ACMTrainingSetup}
\end{algorithm}

We rely on the update procedure associated with the MAPPO algorithm \cite{PPO}. In this algorithm, the advantage estimate is computed using Generalized Advantage Estimation (GAE) method \cite{gae}, with advantage normalization and value clipping over the complete trajectory. The value function is clipped to avoid large updates, ensuring robust learning. A random mini-batch is sampled from the data buffer, and for each data chunk in the mini-batch, the RNN hidden states for both the policy and the critic are initialized from the first hidden state in the data chunk. The policy and critic parameters are then updated using the Adam optimizer \cite{adam}, based on the computed loss, which incorporates the policy objective, value loss, and entropy regularization for exploration.

\section{Experimental Results}
\label{exp_result}

\subsection{Experimental Setting}

 We conduct experiments in two environments, Cooperative Navigation and Rover-Tower, specifically chosen to emulate real-world scenarios with communication constraints. These two environments replicate communications encountered in space missions and planetary rover exploration. 
 
The Cooperative Navigation environment was chosen because it simulates the space environment, a setting where communications are constrained and sporadic but where efficient coordination is important. We use the Cooperative Navigation environment to evaluate the amount of communication used in each approach. In particular, we aim to study how frequently our approach requires communications between agents compared to other baselines. 

 The Rover-Tower environment was chosen as it simulates a real-world planetary exploration scenario involving a rover and a more capable observing agent. We use the Rover-Tower environment to assess each method's ability to adapt to more complex communication scenarios. In particular, we aim to study how generalizable each method is to different communication scenarios involving agents with different observation and communication abilities. 
 
Both the Cooperative Navigation and Rover-Tower environments are implemented in the multi-agent particle environment (MPE)  framework introduced by \cite{lowe2020multiagent}. MPE involves a set of agents that have a discrete action space where it can apply a control unit acceleration in the $x-$ and $y-$ directions. We evaluate our proposed model in the following two environments:
 
\noindent\textbf{Cooperative Navigation:}  \citet{dolan2023satellite} propose a modified MPE environment that uses the satellite dynamics following the Chlohessy-Wiltshire model \cite{vallado_mcclain_2007}, where a set of $n$ satellites are moving within a 2D plane. Each satellite is tasked with rendezvousing to an assigned goal location by the end of the episode. The Chlohessy-Wiltshire equations are a set of second-order differential equations whose $x$ and $y$ components are coupled, thereby increasing the difficulty of the simulation dynamics. We use this modified environment to test our experiment due to its relevancy to the problems we are interested in (e.g., environments with limited communications) and because the agents involved will continue along their relative trajectories even when not communicating with one another. In the Cooperative Navigation task, satellites are initialized into an environment defined by the world size (e.g., 2 km). We set the initialization such that each satellite is at least  500 meters away from all other satellites and their intended goals. This constraint ensures that satellites have sufficient space to maneuver, preventing the occurrence of deadlock. For each episode, each satellite is reinitialized at a different location with a new goal, ensuring a variety of maneuvers and enhancing the generalizability to different navigation interactions.\\
\textbf{Rover-Tower:} \citet{aac} develop the Rover-Tower task environment where randomly paired agents communicate information and coordinate. The environment consists of 4 "rovers" and 4 "towers", with each episode pairing rovers and towers randomly. The performance of the pair is penalized on the basis of the distance of the rover to its goal. The task is designed to simulate scientific navigation on another planet where there is limited infrastructure and low visibility. Rovers cannot observe their surroundings and must rely on communication from the towers, which can locate rovers and their destination and can send one of 5 discrete communications to their paired rover. In this scenario, communication is highly restricted. Extended descriptions of both environments are included in the appendix.

\subsection{Evaluation Metrics}
We rely on several evaluation metrics to characterize the performance of our algorithm against other baselines. We selected these metrics to assess the agent's ability to successfully navigate to their goals without collisions.
 \begin{itemize}
     \item \textbf{Communication frequency $(f_{\mathrm{comm}})$}: Average ratio of the number of messages passed between agents over the maximum number of communication opportunities in the episode. $f_{\mathrm{comm}}$ of 1 indicates each agent messaged every other agent at every time step in the episode (lower value indicates the model is more message efficient).
     \item \textbf{Success Rate $(S\%)$}: Percentage of episodes in which all $n$ agents are able to get to their goals (higher is better).
     \item \textbf{Fraction of Episode Completed $(T)$}: The fraction of an episode that agents take on average to get to the goal. If the agents do not reach the goal, then the fraction is set to 1. For each episode, the $n$ episode fractions are averaged together; this number is then normalized across all evaluation episodes (lower is better).
     \item  \textbf{Average Collisions} $(\#col)$: the total number of collisions that agents had in an episode, normalized by the number of agents and then normalized by the number of evaluation episodes (lower is better).
 \end{itemize}

While the standard reported metric for MPE environments is the global reward, we have not included it in our example results. Since we introduced a new reward formulation for our model, we found reward comparisons challenging to both interpret and compare, given the different reward functions that the baselines were designed with. Similar to \cite{transfQmix}, we use the success rate metric to indicate the overall success of the total number of agents. 
 
\label{asynch_results_total}

\subsection{Training Details}
In the simulation, every policy is trained with 2M steps over 5 random seeds. All results are averaged over 100 testing episodes. Associated hyperparameters for each baseline are included in the appendix. All models are generated by running on a cluster of computer nodes on a Linux operating system. We use Intel Xeon Gold 6248 processors with 384GB of RAM.

\subsection{Comparison of \algoname with Other Methods}
\begin{table*}[t!]
\centering
\resizebox{\linewidth}{!}{%
\begin{tabular}{|l||c|c|c|c||c|c|c|c||c|c|r|c||c|c|c|c|}
\hline

\multirow{2}{*}{Algorithm}  &\multicolumn{4}{c||}{$N=3$} &  \multicolumn{4}{c||}{$N=5$} & \multicolumn{4}{c||}{$N=7$} & \multicolumn{4}{c|}{$N=10$}\\
\cline{2-17}
 & $f_{\mathrm{comm}}$ $\downarrow$ & {$T \downarrow$} & {\# col $\downarrow$} & {$S\% \uparrow$} &  $f_{\mathrm{comm}}$ $\downarrow$  & {$T \downarrow$} & {\# col $\downarrow$} & $S\% \uparrow$ & $f_{\mathrm{comm}}$ $\downarrow$  &  {$T \downarrow$} & {\# col $\downarrow$} & $S\% \uparrow$&  $f_{\mathrm{comm}}$ $\downarrow$ & {$T \downarrow$} & {\# col $\downarrow$} & $S\% \uparrow$\\    \hline \hline
GCS \cite{gcs}          & 1.0   &    0.36 & 0.34 &  100& 1.0&0.42     &  1.72  &  98     & 1.0 &  0.39   &  2.86        &  100   &1.0 & 0.78  & 7.38  & 1   \\ \hline

 asyncMAPPO \cite{asyncmappo}  & 0.21 & 0.10 & 0.86 & 100& 0.20 &  0.32 &  6.05 &  100   & 0.19 & 0.23  & 12.3 & 100 & 0.15 & 0.14& 25.68 & 100       \\ \hline

Actor-Attention-Critic \cite{aac}         & 0.21  & 0.42  & 0.30 & 100  &0.16&  0.47   &  1.20  &  100   & 0.11&   0.49 &    2.52  & 100 & 0.08 & 0.52  & 4.2  & 100    \\ \hline

TransfQmix \cite{transfQmix}          & 0.13 &     0.83 & 0.02  & 42  & 0.16 & 0.96   & 0.12   &   39  & 0.17 & 0.96   & 0.28     & 33 &0.18 &  0.96  & 0.12 &19    \\ \hline

CACOM \cite{CACOM}    & 0.26   & 0.99 & 0.17 & 0 & 0.12&  0.97 & 0.35   & 0  & 0.12&  0.97 & 0.87& 0 & 0.10&  0.98& 1.46 & 0 \\ \hline
DGN \cite{DGN}       &   0.20   & 0.96 & 0.12 & 0 & 0.13 &0.99              &  0.06  &  0  & 0.09 & 0.99    &  0.07 & 0 & 0.06 & 0.98& 0.42 & 0     \\ \hline \hline
\textbf{\algonameNoSpace}  & \textbf{0.10}   &     \textbf{0.24}  & \textbf{0.45} & \textbf{ 97}&   \textbf{0.08} & \textbf{0.23}   & \textbf{0.85} &  \textbf{98}    & \textbf{0.08} & \textbf{0.25}       &  \textbf{2.16}     &  \textbf{86}  & \textbf{0.05} & \textbf{0.34} & \textbf{6.38}   & \textbf{86} 
\\ \hline
\end{tabular}
}
\caption{Comparison of \algoname with other baseline methods for scenarios with 3, 5, 7, and 10 agents in the Cooperative Navigation environment.     }
\label{table:scalecomparison}
\end{table*}
In this section, we demonstrate that \algoname can effectively learn policies for navigation even in settings with less frequent and asynchronous communications. 

We compare our methodology against several alternative MARL frameworks that seek to provide limited communication. 
\begin{itemize}
\item \textbf{GCS} \cite{gcs}: 
The authors factorize the joint team policy into a graph generator and graph-based coordinated policy to enable coordinated behavior among agents.
\item \textbf{Actor-Attention-Critic} \cite{aac}:  Actor-Attention-Critic uses a centralized critic with an attention mechanism that dynamically selects which agents to attend to at any time point during training.
 \item \textbf{asyncMAPPO} \cite{asyncmappo}: The authors extend multi-agent proximal policy optimization \cite{MAPPO} to the asynchronous setting and apply an invariant CNN-based policy to address intra-agent communication. 
\item \textbf{TransfQMix} \cite{transfQmix}: TransfQMix relies on graph transformers to learn encodings over the state of observable entities, and then a multi-layer perceptron to project the q-values of the actions sampled by the individual agents over the q-value of the joint sampled action.  
\item \textbf{CACOM} \cite{CACOM}: The authors employ a two-stage communication scheme where agents first exchange coarse representations and then use attention mechanisms and learned step size quantization techniques to provide personalized messages for the receivers. 
\item \textbf{DGN} \cite{DGN}: DGN relies on graph convolutional networks to model the relational representation and implicitly models the action coordination. 
\end{itemize}
\subsubsection{Performance on Cooperative Navigation Environment}
Table \ref{table:scalecomparison} compares the performance of \algoname against the other baselines. Although having a small number of collisions is better, the policies of some of the baseline algorithms do not significantly move the agents from their initial positions after training and, hence, do not get to the goal. This leads to them having a lower number of collisions. Hence, this metric should be judged by the success rate in context. Similarly, the episode completion rate and communication frequency should be considered within the context of the overall success and collision rates.

When evaluating \algonameNoSpace's performance in the context of these other baselines, our method is able to achieve high success rates and relatively low collision rates, despite 26$\%$ fewer messages being passed between agents. The temporal graph formulation of our model, which inherently allows communications to be masked to reduce communication overhead during training, leads to a method capable of handling trade-offs between communication frequency, success, and collision avoidance. 

When comparing \algoname against other baselines, there are immediate takeaways from the $n=3$ agent case. GCS relies on an acyclic uni-directional graph representation that requires the most recent action selection at the step prior, resulting in high success rates at the cost of a significantly higher communication frequency.  Both asyncMAPPO and Actor-Attention-Critic demonstrate comparable performance in success and collision rates for $n=3$ agents. Similar to the design of \algonameNoSpace, Actor-Attention-Critic is designed to dynamically select which agents to focus on. This reduces $f_\mathrm{comm}$ and leads to improved success and collision rates. However, this attention mechanism overlooks relationships between agents captured by the graph representation used in \algonameNoSpace, leading Actor-Attention-Critic to have a higher communication frequency and episode completion rates. 

The performance of TransfQmix is comparatively less effective. TransfQmix exhibits the lowest collision rates of the algorithms evaluated but at the cost of a low success rate. As stated previously, low collision rates should be considered in the context of the success rate and episode completion rate, as it is possible for agents to learn a policy in which they do not move at all.  

Both DGN and CACOM fail to learn meaningful synthesis of agent communication and have poor success rates as a result. DGN relies on a graph convolutional network, where all neighboring agents contribute equally to the aggregation of node features. The poor performance of DGN in this setting suggests that equal weighting of nearby agents leaves agents unable to capture nuances in the graph structure that are captured through our agent-entity graph embeddings. As noted in CACOM, the learned gate-pruning contains relatively high variance in the Cooperative Navigation environment and is subjected to instability. We believe the complexity of the dynamics in our setting, coupled with the asynchronous formulation, resulted in learning instability in the gating function and, subsequently, poor performance.

 When considering larger numbers of agents ($n=5, 7, 10$), we see similar trends. CACOM and DGN continue to struggle to meaningfully encode information from nearby neighbors at scale.  While asyncMAPPO maintains its strong performance (as evidenced by its success rate), it also possesses a significantly higher number of collisions than \algonameNoSpace. The performance of Actor-Attention-Critic and \algoname are similar. However, the Actor-Attention critic approach requires more frequent communication between agents and results in more collisions in the $n=5$ and $n=7$ cases. As the number of agents increases to $n=10$, we note that the performance of GCS suffers. Upon further inspection, we found that GCS could match the performance it had on the $n=3,5,7$ cases with additional training time (e.g., 2 million steps vs. 5 million steps).  This decrease in performance in GCS suggests that the fully connected graph feature of this model serves to increase computational training time and hinders the ease of scaling the model.   

We note that with the increased number of agents, the communication frequency of all algorithms generally decreases. This can be attributed to the increase in world size, resulting in a less dense environment and fewer communications relative to the number of total communications that would be possible for $n=10$ agents.  As a result, for the $n=10$ case, the communication frequencies are relatively low. 
\subsubsection{Performance on Rover-Tower Environment}
Table \ref{rover-tower_guy} shows \algoname against the best-performing baselines from the prior experiment. As a reminder, the reward function associated with this environment does not include any collision penalty, so we do not include the $\#col$ metric. 
 \begin{table}[]
\centering
\begin{tabular}{|l||c|c|c|}
\hline
\multirow{3}{*}{Algorithm} & \multicolumn{3}{c|}{Metrics}  \\

\cline{2-4}

 & $f_{\mathrm{comm}}$ $ \downarrow$ & {$T \downarrow$}  & {$S\% \uparrow$} \\  
 
 \hline \hline
Actor-Attention-Critic \cite{aac}         &   0.21  & 0.84 &     56\%  \\ \hline
AsyncMAPPO \cite{asyncmappo}       & 0.24  & 0.98  & 0\%       \\ \hline
TransfQmix \cite{transfQmix}    & 0.40 &  0.98  & 0\% \\ \hline
\textbf{\algonameNoSpace} (our method)     &\textbf{ 0.14} & \textbf{0.55} &     \textbf{50\%} 
\\ \hline
\end{tabular}
\vspace{2mm}
\caption{Comparison of \algoname with other baseline methods for scenarios with 4 rovers and 4 towers in the Rover-Tower environment.     }
\label{rover-tower_guy}
\end{table}
In this environment, rovers must rely on encoded messages from their corresponding tower to determine their action selection, whereas towers have more advanced observation abilities. To account for these two classes, Actor-Attention-Critic creates a separate network for the rover class and the tower class, whereas \algoname does not. Despite the fact that \algoname is using a singular network to represent both the rovers and the towers, it still achieves a comparable success rate to the Actor-Attention-Critic. Additionally, \algoname relies on less communication and produces faster episode completion rates than other baselines, suggesting that \algoname is a more efficient, generalizable communication protocol for this environment.

\subsubsection{Visualizing the Graph Transformer Weights}
\begin{figure}[htbp!]
\centering   
    \includegraphics[width=10.5cm]{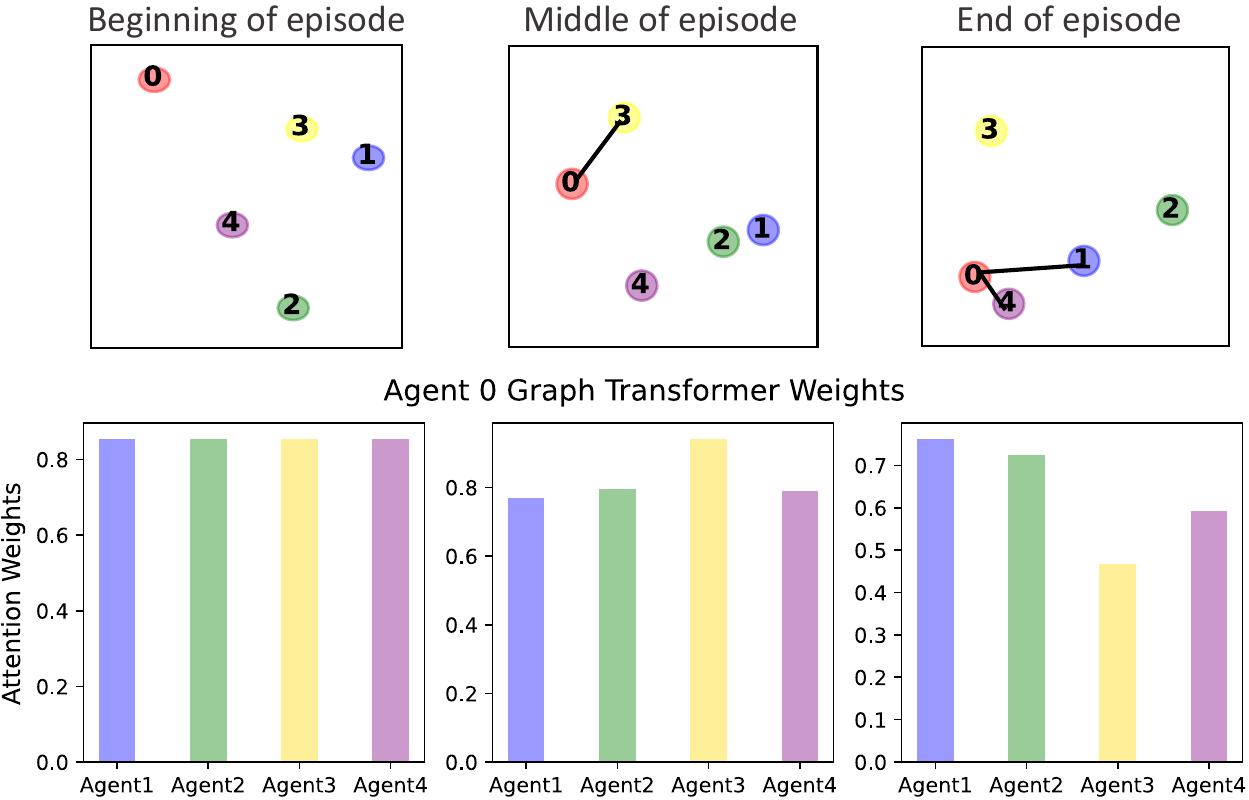}
    \caption{Attention weights for agent 0 in the $n=5$ agent Cooperative Navigation task. We compare the changes in graph transformer attention at three discrete periods during the episode at the beginning, middle, and end. }
    \label{visual_att}
\end{figure}

To better understand the underlying mechanisms of our graph transformer communication protocol, we visualize the graph transformer attention weights at three different times in an episode for a single agent in the Cooperative Navigation environment. In Figure \ref{visual_att}, the leftmost panel corresponds to the weights at time $\tau=0$. In this panel,  agent $0$ is unconnected to any other agents, and thus, the attention weighting for all other nodes is perfectly equal.  The center panel corresponds with $\tau=6$ for agent 0, and at this point in the episode, agent $0$ is now able to communicate with agent $3$ due to their proximity. As a result, the attention weighting of the messages from agent 3 at this time is higher than the other agents, as shown by the attention weighting panel. 
The final rightmost panel corresponds with $\tau=18$ for agent 0. At this point in the episode, the agent is now within the communication range of agents 1 and 4, as shown by the connecting edges in the agent locations panel.  Similar to what we observed in the second panel, the corresponding attention weights to these two agents are also higher. Interestingly, we also find that the attention weight for agent 2 is also fairly large, despite not being connected via graph. Upon further inspection, we found that this is attributed to the communication frequencies of agents 0 and 2; both of these two agents communicated more frequently throughout the episode than agent 0 did with agents 3 and 4.  We find that the graph transformer communication protocol learns to attend to both agents in proximity to the active agent (e.g., panel 2) but also to those agents from whom it gets more frequent communication (e.g., panel 3). Therefore, the model implicitly learns the trade-off between agent proximity and frequency of communication from specific agents.

\subsection{Ablation Studies}
\subsubsection{Impact of Graph Transformer}
 To verify the effectiveness of our graph transformer communication protocol, we conduct an ablation study on the Cooperative Navigation environment. We train and evaluate two models on $n=10$ agents: (1) \algonameNoSpace, our graph transformer-based communication protocol for multi-agent reinforcement learning, and (2) MARL, our stripped-down asynchronous multi-agent reinforcement learning formulation that only communicates when agents are active at the same time step. We compare the differences between the two models in Table \ref{twomodels}. 
\begin{table}[ht]
\centering
\begin{tabular}{|l|| l |c |}
\hline
   \textbf{Max Number  }  & \multicolumn{2}{|c|}{ \textbf{Percent Improvements}}  \\\cline{2-3}

\textbf{of Active Agents }  &  \multirow{1}{*}{\begin{tabular}[c]{@{}c@{}}\# col \end{tabular}}  & \multirow{1}{*}{\begin{tabular}[c]{@{}c@{}} S\% \end{tabular}} 
      \\ \hline \hline
  2 &   92.4\%   &  83.1\%   \\ \hline
 3 &  74.7\%     &  64.2\%   \\ \hline
 5 & 39.7\%     &    69.4\%              \\ \hline
\end{tabular}
\vspace{2mm}
\caption{Comparison of our model against a simplified variant that has no graph transformer communication protocol for $n=10$ agents.}
\label{twomodels}
\end{table}
We aimed to determine whether there was a relationship between an agent's active status and performance. Specifically, we were interested in determining whether the performance of our algorithm could be attributed to large numbers of agents all active at the same time. To that end, we performed a modified experiment where we fixed the maximum number of agents that could be active at the same time. Then, we compared the difference in performance with and without our graph transformer protocol. 

We note that across all numbers of active agents evaluated, the graph transformer communication protocol led to improvements in goal-reaching and lowered collision rates. The largest improvements from the graph transformer communication protocol occur at smaller numbers of active agents (e.g., 2). These results suggest that in dynamic graphs, smaller graph structures lead to reduced noise and a stronger abstraction of the essential structure of the environment. 

\subsubsection{Reward Formulation}
In developing \algonameNoSpace, we experimented with several different reward structures to balance trade-offs between individual goal-reaching and collaborative path planning. We compare the following reward formulations.
\begin{itemize}
    \item  \textbf{Repeated Reward}: This is the reward structure used in the InforMARL algorithm~\cite{informarl_icml}. For a single agent, the reward is calculated by Equation \ref{eq:reward}.

In this formulation, the goal-reaching reward, $\mathcal{R}_{goal,t}^{(i)}=5 $, is given to each agent for every step it is at the goal. This means that even when an agent has successfully reached its goal and no longer needs to take any further control actions, it is still receiving a reward. This individual agent reward is combined with the reward received by all other agents to calculate the total reward for that step, $\mathcal{R}_{total}^{(i)}$. We combine the individual rewards of each agent into a sum to encourage collaborative behavior. With this reward structure, the largest possible reward is when all $n$ agents reach their goal.

\item \textbf{Piecewise Reward}: In the piecewise structure, each agent $i$ receives a $\mathcal{R}_{goal,t}^{(i)}$ value of +5 at the first time step that it reaches gets within distance $\delta$ of the goal, $p^{goal}_i$. For every time step after, the goal-reaching value is changed in Equation \ref{eq:reward} to be a smaller value.
\begin{equation}
\begin{aligned}
       \mathcal{R}_{goal,t}^{(i)}&= \begin{cases} 
  +5 & t_g= || s_t^{(i)}- p^{goal}_i || \leq \delta  \\
  +0.5 & t> t_g 
  \end{cases}
\end{aligned}
\label{pcw_rew}
\end{equation}
In our analysis, we found that a larger magnitude of goal-reaching reward could obscure penalties for goal-reaching and collision avoidance for the other agents. By adopting the goal-reaching reward to be smaller after the first instance, the objective of the piecewise reward function is to encourage goal-reaching behavior without obscuring the collision and goal-reaching penalties of the other agents. We relied on the same summation function across all agents to encourage collaboration. 
\item \textbf{Single Goal-Reaching Reward}: In the single goal-reaching reward structure, each agent receives a $\mathcal{R}_{goal,t}^{(i)}$ value of +5 at the first time step that it reaches the goal. For every time step after, it receives no reward.  We use boolean variable $\phi$ to designate that the goal-reaching reward has already been allocated to agent $i$. We investigated this structure to determine if there were any residual benefits to learning if the agents received no rewards after they reached their goals. As in the continuous reward structure, we relied on the same summation function across all agents to encourage collaboration. 
\item \textbf{Single Goal-Reaching Reward + Active Agent Sharing}: We adapt the single goal-reaching reward structure to also consider a more complex sharing function amongst agents. In this reward structure, agents only receive collaborative rewards if they communicate with one another during that timestep $t$.  The purpose of this reward structure was to investigate if there was any impact on learned behaviors when agents received information more recently (and when they could operate synchronously for that given step). 
\end{itemize}
\begin{table}[htbp!]
\centering
\begin{tabular}{|l || c | c| c |}
\hline

\textbf{Reward} & \multicolumn{3}{c|}{\textbf{Metrics}}    \\ \cline{2-4}
\textbf{Structure} &  $T$ &  \# col  & \multicolumn{1}{l|}{$S\% $ } \\ \hline \hline

Repeated     & 0.33   & 4.7   &16\%    \\ \hline
Piecewise    & 0.23 & 1.6   & 45\%    \\ \hline
Single        &  0.31   & 1.21  &41\%     \\ \hline
\textbf{Single + Active}   & \textbf{0.24}  & \textbf{0.45}  & \textbf{97\%}    \\ \hline
\end{tabular}
\vspace{2mm}
\caption{Comparison of the impact of several reward formulations on the resultant performance in the asynchronous setting for $n=3$ agents.}
\label{reward_ablation}
\end{table}
Table \ref{reward_ablation} compares the performance of the different reward structures. Across the four reward structures, we have found that the \textit{single goal-reaching reward + active agent sharing} case produced the best results in terms of success rate and average collision number. 

This result indicates that at an individual level, learning improves in an asynchronous setting when the goal-reaching reward is only received once (as demonstrated by the lower collision rates for the two single-goal-reaching reward columns). By removing the repeated additive goal-reaching reward, the other agents are able to better refine their behaviors and recognize collisions. 

When comparing the single-goal reaching reward and the active-agent sharing case, we have empirically found that the extent of shared collaboration plays an important role in the success rate of the agents. When the agents had a shared reward, this created a lagged reward function, where the reward structure was determined by the last reward produced by one of the agents. By comparison, when active agents share their reward, this creates a reward function that is informed by the rewards produced by the actions taken at that step.

\section{Conclusion} 

We introduced \algonameNoSpace, a graph-transformer communication protocol for asynchronous multi-agent reinforcement learning designed to address the problem of coordination in environments where agents cannot communicate regularly. Each agent's graph transformer utilizes a dynamic, weighted, directed graph to learn a communication protocol with other active agents in its vicinity. First, we showed that our method required less communication between agents and still produced similar success and collision rates as other multi-agent reinforcement learning approaches. Then, we evaluated \algonameNoSpace's performance in the more challenging Rover-Tower environment and found that our framework produces comparable results to other methods that require a separate network for the two agent classes. We further examined the workings of our graph transformer mechanism over the course of an episode and found that it effectively balances the trade-offs between the proximity of other agents and their active status.

Through ablation studies, we demonstrated the effectiveness of our graph transformer-based communication protocol, as well as the importance of reward structures in asynchronous settings. In future research, we aim to explore more advanced communication protocol architectures that can model different action-communication constraints common in real-world settings. Additionally, we want to investigate the feasibility of integrating additional mechanisms like control barrier functions to reduce the overall number of collisions.

\section{Acknowledgements}

The authors would like to thank the MIT SuperCloud \citep{supercloud} and the Lincoln Laboratory Supercomputing Center for providing high-performance computing resources that have contributed to the research results reported in this paper. This work was supported in part by NASA under grant \#80NSSC23M0220 and the University Leadership Initiative (grant \#80NSSC20M0163), but this article solely reflects the opinions and conclusions of its authors and not any NASA entity. The research was sponsored by the Department of the Air Force Artificial Intelligence Accelerator and was accomplished under Cooperative Agreement Number FA8750-19-2-1000. The views and conclusions contained in this document are those of the authors and should not be interpreted as representing the official policies, either expressed or implied, of the United States Air Force or the U.S. Government. The U.S. Government is authorized to reproduce and distribute reprints for Government purposes notwithstanding any copyright notion herein. Sydney Dolan was supported in part by the National Science Foundation Graduate Research Fellowship under Grant No. 1650114. J. Aloor was also supported in part by a Mathworks Fellowship.

\bibliographystyle{unsrtnat}
\bibliography{main} 

\newpage
\section{Appendix}

\subsection{Baseline Implementation Details}
We rely on the following implementations for each baseline and provide links to those implementations here. Note that we used the same hyperparameters as used in their original implementations, assuming that they were optimal.
\begin{enumerate}
    \item asyncMAPPO: \url{https://github.com/yang-xy20/async_mappo/tree/main}
    \item GCS: \url{https://github.com/LXXXXR/GCS_aamas337/tree/master}
    \item DGN: \url{https://github.com/jiechuanjiang/pytorch_DGN}
    \item CACOM: \url{https://github.com/LXXXXR/CACOM/tree/main}
    \item Actor-Attention Critic: \url{https://github.com/shariqiqbal2810/MAAC/tree/master}
    \item TransfQmix: \url{https://github.com/mttga/pymarl_transformers/tree/main}
\end{enumerate}

\subsection{Environment Implementation Details}
We rely on the following implementations for the two environments we used in our experiments.
\begin{enumerate}
    \item Cooperative Navigation: \url{https://github.com/sydneyid/satellite-cooperative-nav}
    \item Rover-Tower: \url{https://github.com/shariqiqbal2810/MAAC/tree/master}
\end{enumerate}

\subsection{Cooperative Navigation Environment Description}
There are $n$ agents and $n$ goals, along with static obstacles in the environment. Each agent is supposed to go to its distinct goal while avoiding collisions with other entities in the environment. Agents start at random locations at the beginning of each episode; the corresponding goals are also randomly distributed. Agents are governed by Clohessy-Wiltshire Equations \cite{vallado_mcclain_2007}:
    \begin{align}
    \ddot{x} - 3n^2x-2n\dot{y} &= u_x \label{xdotchw} \\
    \ddot{y}  +2n\dot{x} &= u_y \label{ydotchw} \\
    \ddot{z}  + n^2z &= u_z\label{zdotchw}
    \end{align}
The above equations consider a localized coordinate system centered around one of the satellites, referred to as the \emph{target}). The target satellite is assumed to have a circular orbit with an orbital rate of $\omega_n$. The coordinates are defined such that $x$ is measured radially outward from the target, $y$ is along the orbit track of the target body, and $z$ is along the angular momentum. $u_x$, $u_y$, and $u_z$ represent the acceleration in the x,y, and z- directions. A full derivation of the perturbed equations can be found in \cite{cw_j2}. While there exist additional environmental perturbations, such as solar radiative pressure or three-body effects, their impacts are magnitudes smaller \citep{walter_2018}. In our method, we only use the $x$ and $y$ to dictate the agent's motion, as we are assuming all agents act within the same $z$ plane. 
The experimental setup for each result is reported in Table \ref{expset}.

\begin{table}[h]
\centering
\begin{tabular}{|l|l|l|l|}
\hline
\textbf{Num} & \textbf{Num} & \textbf{World} & \textbf{Episode} \\ 
\textbf{Agents} & \textbf{Obstacles} & \textbf{Size} & \textbf{Length} \\ \hline
3                   & 3                      & 2                   & 125                     \\ \hline
5                   & 3                      & 2                   & 125                     \\ \hline
7                   & 3                      & 3                   & 125                     \\ \hline
10                  & 3                      & 3                   & 125                     \\ \hline
\end{tabular}
\caption{Experimental set up for experiments in Results section}
\label{expset}
\end{table}
\newpage
\subsection{Rover-Tower Environment Description}
In the Rover-Tower environment, there are 4 rovers and 4 towers. 
Tower agents can send one of 5 discrete communication messages to their paired rover at each time step. Towers cannot move but can communicate with rovers so that rovers can move towards their corresponding goal. Each pair of rover and tower are negatively rewarded by the distance of the rover to its goal at each episode. In our setup, the communication is integrated into the environment (in the tower's action space and the rovers observation space), rather than being explicitly part of the model.

\subsection{GCS Extended Training Result}
\begin{table}[htbp]
\centering
\begin{tabular}{|l||c||c||c |l |l |}
\hline
\multirow{2}{*}{Algorithm} & Broad                       & Message                  & \multicolumn{3}{c|}{N=10}                                                       \\ \cline{4-6} 
                           & Cast                 & Encoding     & \multicolumn{1}{c|}{T} & \multicolumn{1}{c|}{\# col} & \multicolumn{1}{c|}{S\%} \\ \hline \hline
GCS \cite{gcs}  & Global & GAT & 0.39  &  5.97    &  100  \\ \hline
\end{tabular}
\caption{Results for GCS when the training period is extended to 5,000,000 steps. Evaluations across 100 episodes with $n = 10$ agents.}
\end{table}

\subsection{Associated Key Dependencies}
We rely on the following package versions to support our algorithm:
\begin{enumerate}
     \item gym = 0.26.2
     \item torch = 1.13.1
     \item torch-geometric = 2.3.1
     \item tensorboardX = 2.6.2.2
     \item wandb = 0.17.4
\end{enumerate}

\subsection{Hyperparameters}
\begin{table}[htbp]
  \centering
  \begin{tabular}{lc}
 \hline
    Common Hyperparameters     & Value \\
 \hline
    number of att heads     & 3 \\
    GAT Encoder num heads  & 4\\
    num layers      &  4 \\
    decoder hidden dim     &  64 \\
    \hline
  \end{tabular}
  \caption{Common Hyperparameters used in GCS}
  \label{table:hp_gcs}
\end{table}

\begin{table}[htbp]
  \centering
  \begin{tabular}{lc}
    \hline
    Common Hyperparameters     & Value \\
     \hline
    recurrent data chunk length & 10 \\
    gradient clip norm     & 10.0 \\
    gae lambda     & 0.95 \\
    gamma     & 0.99 \\
    value loss     &  Huber loss \\
    huber delta     &  10.0 \\
    batch size     &  num envs $\times$ buffer length $\times$ num agents\\
    mini batch size     &  batch size / mini-batch\\
    optimizer     &  Adam \\
    optimizer epsilon     &  1e-5 \\
    weight decay     &  0 \\
    network initialisation     &  Orthogonal \\
    use reward normalisation     &  True \\
    use feature normalisation     &  True \\
    num envs & 64 \\
    buffer length & 125 \\
    \hline
  \end{tabular}
  \caption{Common Hyperparameters used in asyncMAPPO, GCS, and AsyncCoMARL }
  \label{table:mappo-hp}
\end{table}

\begin{table}[htbp]
  \centering
  \begin{tabular}{lc}
     \hline
    Common Hyperparameters     & Value \\  \hline
    attention dim     & 32 \\
    hidden layer size     &  64 \\
    mi loss weight    &  0.001 \\
    entropy loss weight     &  0.01\\
    encoder dimension & 8\\
    request dimension & 10\\
    response dimension & 20 \\
    3 Agent obs segments & $ \langle 1\times 4, 4 \times 2, 1 \times 6 \rangle$ \\ \hline
  \end{tabular}
  \caption{Common Hyperparameters used in CACOM}
  \label{table:hp_cacpm}
\end{table}

\end{document}